\documentclass[%
 reprint,
 amsmath,amssymb,
 aps,
]{revtex4-1}

\usepackage{graphicx}
\usepackage{dcolumn}
\usepackage{bm}
\usepackage{lipsum}
\usepackage{verbatim}

\begin{document}


\title{Training of photonic neural networks through \textit{in situ} backpropagation}

\author{Tyler W. Hughes, Momchil Minkov, Yu Shi, Shanhui Fan}
\affiliation{Ginzton Laboratory, Stanford University, Stanford, CA, 94305.}


\date{\today}

\begin{abstract}
Recently, integrated optics has gained interest as a hardware platform for implementing machine learning algorithms.  Of particular interest are artificial neural networks, since matrix-vector multiplications, which are used heavily in artificial neural networks, can be done efficiently in photonic circuits. The training of an artificial neural network is a crucial step in its application. However, currently on the integrated photonics platform there is no efficient protocol for the training of these networks. In this work, we introduce a method that enables highly efficient, \textit{in situ} training of a photonic neural network. We use adjoint variable methods to derive the photonic analogue of the backpropagation algorithm, which is the standard method for computing gradients of conventional neural networks. We further show how these gradients may be obtained exactly by performing intensity measurements within the device. As an application, we demonstrate the training of a numerically simulated photonic artificial neural network. Beyond the training of photonic machine learning implementations, our method may also be of broad interest to experimental sensitivity analysis of photonic systems and the optimization of reconfigurable optics platforms.
\end{abstract}

\maketitle

\section{Introduction} \label{sec:intro}

Artificial neural networks (ANNs), and machine learning in general, are becoming ubiquitous for an impressively large number of applications \cite{LeCun2015}. This has brought ANNs into the focus of research in not only computer science, but also electrical engineering, with hardware specifically suited to perform neural network operations actively being developed. There are significant efforts in constructing artificial neural network architectures using various electronic solid-state platforms \cite{Merolla2014,Prezioso2015}, but ever since the conception of ANNs, a hardware implementation using optical signals has also been considered \cite{Abu-Mostafa1987, Jutamulia1996}. In this domain, some of the recent work has been devoted to photonic spike processing \cite{Rosenbluth2009, Tait2014} and photonic reservoir computing \cite{Brunner2013, Vandoorne2014}, as well as to devising universal, chip-integrated photonic platforms that can implement any arbitrary ANN \cite{Shainline2017, Shen2017}.  Photonic implementations benefit from the fact that, due to the non-interacting nature of photons, linear operations -- like the repeated matrix multiplications found in every neural network algorithm -- can be performed in parallel, and at a lower energy cost, when using light as opposed to electrons. 

A key requirement for the utility of any ANN platform is the ability to train the network using algorithms such as error backpropagation \cite{Rumelhart1986}. Such training typically demands significant computational time and resources and it is generally desirable for error backpropagation to implemented on the same platform. This is indeed possible for the technologies of Refs. \cite{Merolla2014, Graves2016, hermans2015trainable} and has also been demonstrated e.g. in memristive devices \cite{alibart2013pattern, Prezioso2015}. In optics, as early as three decades ago, an adaptive platform that could approximately implement the backpropagation algorithm experimentally was proposed \cite{wagner1987multilayer,psaltis1988adaptive}. However, this algorithm requires a number of complex optical operations that are difficult to implement, particularly in integrated optics platforms. Thus, the current implementation of a photonic neural network using integrated optics has been trained using a model of the system simulated on a regular computer \cite{Shen2017}. This is inefficient for two reasons. First, this strategy depends entirely on the accuracy of the model representation of the physical system. Second, unless one is interested in deploying a large number of identical, fixed copies of the ANN, any advantage in speed or energy associated with using the photonic circuit is lost if the training must be done on a regular computer.  Alternatively, training using a brute force, \textit{in situ} computation of the gradient of the objective function has been proposed \cite{Shen2017}. However, this strategy involves sequentially perturbing each individual parameter of the circuit, which is highly inefficient for large systems.

In this work, we propose a procedure, which we label the time-reversal interference method (TRIM), to compute the gradient of the cost function of a photonic ANN by use of only \textit{in situ} intensity measurements.  Our procedure works by \textit{physically} implementing the adjoint variable method (AVM), a technique that has typically been implemented computationally in the optimization and inverse design of photonic structures \cite{Georgieva2002, Veronis2004, hughes2017method}.  Furthermore, the method scales in constant time with respect to the number of parameters, which allows for backpropagation to be efficiently implemented in a hybrid opto-electronic network.  Although we focus our discussion on a particular hardware implementation of a photonic ANN, our conclusions are derived starting from Maxwell’s equations, and may therefore be extended to other photonic platforms.

The paper is organized as follows:  In Section \ref{sec:ANN}, we introduce the working principles of a photonic ANN based on the hardware platform introduced in Ref. \cite{Shen2017}. We also derive the mathematics of the forward and backward propagation steps and show that the gradient computation needed for training can be expressed as a modal overlap.  Then, in Section \ref{sec:adjoint} we discuss how the adjoint method may be used to describe the gradient of the ANN cost function in terms of physical parameters.  In Section \ref{sec:experimental}, we describe our procedure for determining this gradient information experimentally using \textit{in situ} intensity measurements.  We give a numerical validation of these findings in Section \ref{sec:numerical} and demonstrate our method by training a model of a photonic ANN in Section \ref{sec:ANN_simulation}. We provide final comments and conclude in Section \ref{sec:conclusion}.

\section{The Photonic Neural Network}  \label{sec:ANN}
\label{sec:ann}

In this Section, we introduce the operation and gradient computation of a feed-forward photonic ANN.  In its most general case, a feed-forward ANN maps an input vector to an output vector via an alternating sequence of linear operations and element-wise nonlinear functions of the vectors, also called `activations'.  A cost function, $\mathcal{L}$, is defined over the outputs of the ANN and the matrix elements involved in the linear operations are tuned to minimize $\mathcal{L}$ over a number of training examples via gradient-based optimization.  The `backpropagation algorithm' is typically used to compute these gradients analytically by sequentially utilizing the chain rule from the output layer backwards to the input layer.

\begin{figure*}
\centering
\includegraphics[width=0.7\textwidth]{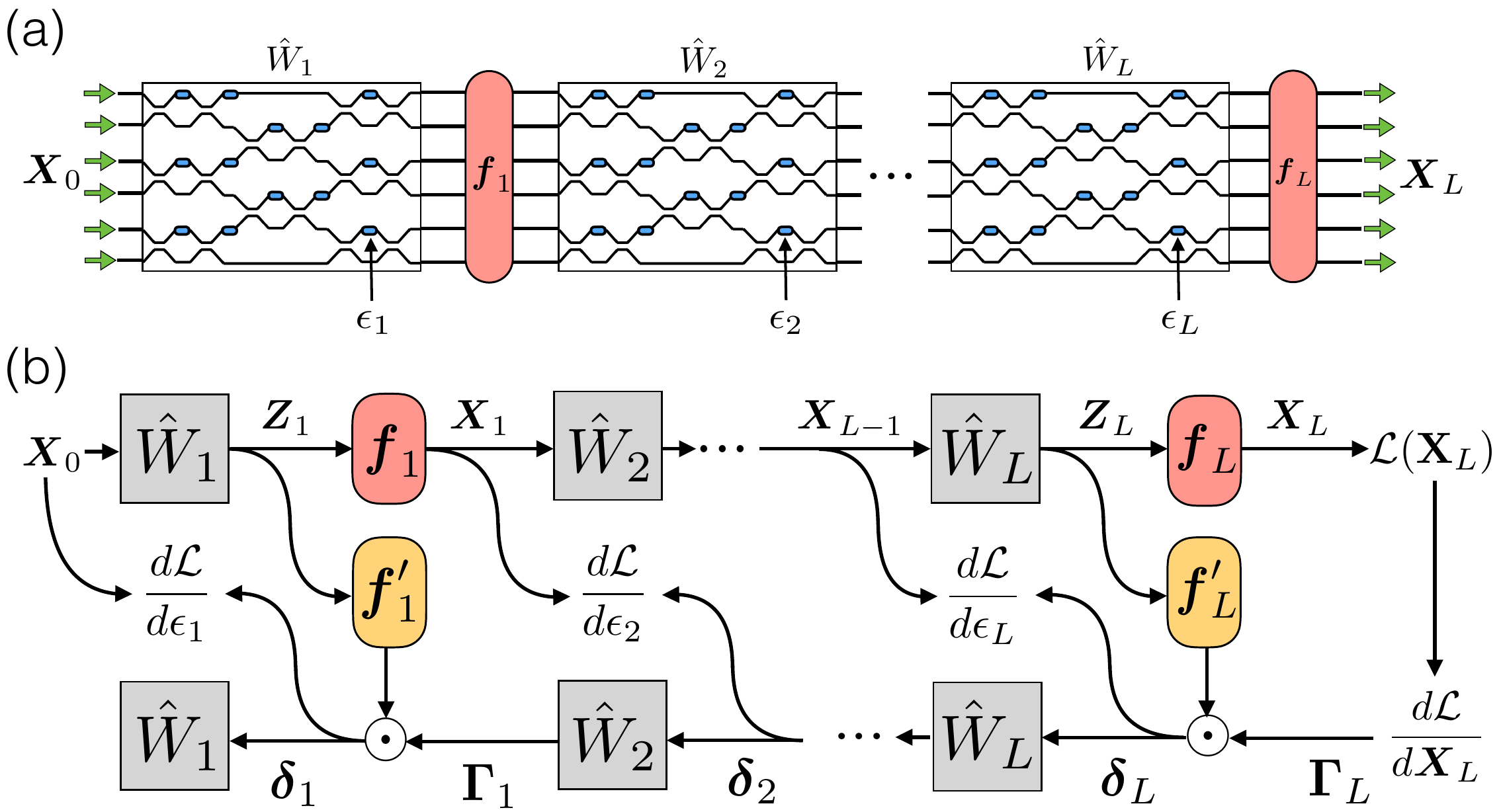}
\caption{\label{fig:backprop} (a) A schematic of the ANN architecture demonstrated in Ref \cite{Shen2017}.  The boxed regions correspond to OIUs that perform a linear operation represented by the matrix $\hat{W}_l$.  Integrated phase shifters (blue) are used to control the OIU and train the network.  The red regions correspond to nonlinear activations $\mathbf{f}_l(\cdot)$.  (b) Illustration of operation and gradient computation in an ANN.  The top and bottom rows correspond to the forward and backward propagation steps, respectively.  Propagation through a square cell corresponds to matrix multiplication.  Propagation through a rounded region corresponds to activation.  $\odot$ is element-wise vector multiplication.}
\end{figure*}

Here, we will outline these steps mathematically for a single training example, with the procedure diagrammed in Fig. \ref{fig:backprop}a.  We focus our discussion on the photonic hardware platform presented in \cite{Shen2017}, which performs the linear operations using optical interference units (OIUs).  The OIU is a mesh of controllable Mach-Zehnder interferometers (MZIs) integrated in a silicon photonic circuit. By tuning the phase shifters integrated in the MZIs, any unitary $N \times N$ operation on the input can be implemented \cite{Reck1994,Clements2016}, which finds applications both in classical and quantum photonics \cite{Carolan2015, Harris2017}.  In the photonic ANN implementation from Ref. \cite{Shen2017}, an OIU is used for each linear matrix-vector multiplication, whereas the nonlinear activations are performed using an electronic circuit, which involves measuring the optical state before activation, performing the nonlinear activation function on an electronic circuit such as a digital computer, and preparing the resulting optical state to be injected to the next stage of the ANN.

We first introduce the notation used to describe the OIU, which consists of a number, $N$, of single-mode waveguide input ports coupled to the same number of single-mode output ports through a linear and lossless device. In principle, the device may also be extended to operate on a different number of inputs and outputs. We further assume directional propagation such that all power flows exclusively from the input ports to the output ports, which is a typical assumption for the devices of Refs. \cite{Miller2013a, Shen2017, Harris2017, Carolan2015, Reck1994,Miller2013,Clements2016}. In its most general form, the device implements the linear operation
\begin{equation}
\hat{W}\mathbf{X}_\textrm{in} = \mathbf{Z}_\textrm{out},
\label{eq:original_linear_system}
\end{equation}
where $\mathbf{X}_\textrm{in}$ and $\mathbf{Z}_\textrm{out}$ are the modal amplitudes at the input and output ports, respectively, and $\hat{W}$, which we will refer to as the transfer matrix, is the off-diagonal block of the system's full scattering matrix,
\begin{equation}
\begin{pmatrix}
\mathbf{X}_\textrm{out} \\
\mathbf{Z}_\textrm{out}
\end{pmatrix} = \begin{pmatrix}
0 & \hat{W}^T \\
\hat{W} & 0
\end{pmatrix}  
\begin{pmatrix}
\mathbf{X}_\textrm{in} \\
\mathbf{Z}_\textrm{in}
\end{pmatrix}.
\label{eq:smatrix}
\end{equation}
Here, the diagonal blocks are zero because we assume forward-only propagation, while the off-diagonal blocks are the transpose of each other because we assume a reciprocal system. $\mathbf{Z}_\textrm{in}$ and $\mathbf{X}_\textrm{out}$ correspond to the input and output modal amplitudes, respectively, if we were to run this device in reverse, i.e. sending a signal in from the output ports.


Now we may use this notation to describe the forward and backward propagation steps in a photonic ANN.  In the forward propagation step, we start with an initial input to the system, $\mathbf{X}_0$, and perform a linear operation on this input using an OIU represented by the matrix $\hat{W}_{1}$.  This is followed by the application of a element-wise nonlinear activation, $\mathbf{f}_1(\cdot)$, on the outputs, giving the input to the next layer.  This process repeats for the each layer $l$ until the output layer, $L$.  Written compactly, for $l = 1~...~L$
\begin{equation}
\mathbf{X}_l = \mathbf{f}_l(\hat{W}_l\mathbf{X}_{l-1}) \equiv \mathbf{f}_l(\mathbf{Z}_l).
\label{eq:NN_recursive}
\end{equation}
Finally, our cost function $\mathcal{L}$ is an explicit function of the outputs from the last layer, $\mathcal{L} = \mathcal{L}(\mathbf{X}_L)$.  This process is shown in Fig. \ref{fig:backprop}(a).

To train the network, we must minimize this cost function with respect to the linear operators, $\hat{W}_l$, which may be adjusted by tuning the integrated phase shifters within the OIUs.  While a number of recent papers have clarified how an individual OIU can be tuned by sequential,  \textit{in situ} methods to perform an arbitrary, pre-defined operation \cite{Miller2013, Miller2013a, Miller2015, Annoni2017}, these strategies do not straightforwardly apply to the training of ANNs, where nonlinearities and several layers of computation are present.  In particular, the training of ANN requires gradient information which is not provided directly in the methods of Ref. \cite{Miller2013, Miller2013a, Miller2015, Annoni2017}.

In Ref. \cite{Shen2017}, the training of the ANN was done \textit{ex situ} on a computer model of the system, which was used to find the optimal weight matrices $\hat{W}_l$ for a given cost function. Then, the final weights were recreated in the physical device, using an idealized model that relates the matrix elements to the phase shifters. Ref. \cite{Shen2017} also discusses a possible \textit{in situ} method for computing the gradient of the ANN cost function through a serial perturbation of every individual phase shifter (`brute force' gradient computation). However, this gradient computation has an unavoidable linear scaling with the number of parameters of the system.  The training method that we propose here operates without resorting to an external model of the system, while allowing for the tuning of each parameter to be done in parallel, therefore scaling significantly better with respect to the number of parameters when compared to the brute force gradient computation.

To introduce our training method we first use the backpropagation algorithm to derive an expression for the gradient of the cost function with respect to the permittivities of the phase shifters in the OIUs. In the following, we denote $\epsilon_l$ as the permittivity of a single, arbitrarily chosen phase shifter in layer $l$, as the same derivation holds for each of the phase shifters present in that layer.  Note that $\hat{W}_l$ has an explicit dependence on $\epsilon_l$, but all field components in the subsequent layers also depend implicitly on $\epsilon_l$.

As a demonstration, we take a mean squared cost function
\begin{align}
\mathcal{L} &= \frac{1}{2}\big(\mathbf{X}_L -\mathbf{T} \big)^\dagger \big( \mathbf{X}_L -\mathbf{T} \big),
\label{eq:NN_forward}
\end{align}
where $\mathbf{T}$ is a complex-valued target vector corresponding to the desired output of our system given input $\mathbf{X}_0$.

Starting from the last layer in the circuit, the derivative of the cost function with respect to the permittivity $\epsilon_L$ of one of the phase shifters in the last layer is given by

\begin{align}
\frac{d\mathcal{L}}{d\epsilon_L} &= \mathcal{R}\left\{\big(\mathbf{X}_L - \mathbf{T} \big)^\dagger \frac{d\mathbf{X}_L}{d\epsilon_L} \right\}\\
    &= \mathcal{R}\left\{ \left(  \bm{\Gamma}_L \odot {\mathbf{f}_L}^{'}(\mathbf{Z}_{L}) \right)^T \frac{d \hat{W}_L}{d\epsilon_L}\mathbf{X}_{L-1} \right\}\\ 
    &\equiv \mathcal{R}\left\{ \boldsymbol{\delta}_L^T \frac{d \hat{W}_L}{d\epsilon_L} \mathbf{X}_{L-1} \right\},
\label{eq:backprop_L}
\end{align}
where $\odot$ is element-wise vector multiplication, defined such that, for vectors $\mathbf{a}$ and $\mathbf{b}$, the $i$-th element of the vector $\mathbf{a} \odot \mathbf{b}$ is given by $a_i b_i$. $\mathcal{R}\{\cdot\}$ gives the real part, ${\mathbf{f}_l}^{'}(\cdot)$ is the derivative of the $l$th layer activation function with respect to its (complex) argument.  We define the vector $\bm{\delta}_L \equiv \bm{\Gamma}_L \odot {\mathbf{f}_L}^{\,'}$ in terms of the error vector $\bm{\Gamma}_L \equiv \big(\mathbf{X}_L - \mathbf{T} \big)^*$.

For any layer $l < L$, we may use the chain rule to perform a recursive calculation of the gradients
\begin{align}
\bm{\Gamma}_l &= \hat{W}^T_{l+1} \bm{\delta}_{l+1} \label{eq:backprop_general_Gamma}
\\
\bm{\delta}_l &= \bm{\Gamma}_l \odot {\mathbf{f}_l}^{'}(\mathbf{Z}_{l}) \label{eq:backprop_general2}
\\
\frac{d\mathcal{L}}{d\epsilon_l} &= \mathcal{R}\left\{ \bm{\delta}_l^T \frac{d \hat{W}_l}{d\epsilon_l} \mathbf{X}_{l-1} \right\}.
\label{eq:backprop_general}
\end{align}
Figure \ref{fig:backprop}(b) diagrams this process, which computes the $\bm{\delta}_l$ vectors sequentially from the output layer to the input layer.  A treatment for non-holomorphic activations is derived Appendix A.

We note that the computation of $\bm{\delta}_l$ requires performing the operation $\bm{\Gamma}_l = \hat{W}^T_{l+1} \bm{\delta}_{l+1}$, which corresponds physically to sending $\bm{\delta}_{l+1}$ into the output end of the OIU in layer $l+1$.  In this way, our procedure `backpropagates' the vectors $\bm{\delta}_l$ and $\bm{\Gamma}_l$ physically through the entire circuit.


\section{Gradient computation using the adjoint variable method} \label{sec:adjoint}

In the previous Section, we showed that the crucial step in training the ANN is computing gradient terms of the form  $\mathcal{R}\left\{\boldsymbol{\delta}_l^T \frac{d \hat{W}_l}{d\epsilon_l} \mathbf{X}_{l-1}\right\}$, which contain derivatives with respect to the permittivity of the phase shifters in the OIUs. In this Section, we show how this gradient may be expressed as the solution to an electromagnetic adjoint problem.

The OIU used to implement the matrix $\hat{W}_l$, relating the complex mode amplitudes of input and output ports, can be described using first-principles electrodynamics.  This will allow us to compute its gradient with respect to each $\epsilon_l$, as these are the physically adjustable parameters in the system. Assuming a source at frequency $\omega$, at steady state Maxwell's equations take the form 
\begin{equation}
\Big[ \hat{\nabla} \times \hat{\nabla} \times ~ -k_0^2 \hat{\epsilon}_r \Big]\mathbf{e} = -i\omega \mu_0 \mathbf{j},
\label{eq:maxwells_equations_physics}
\end{equation}
which can be written more succinctly as
\begin{equation}
\hat{A}(\epsilon_r) \mathbf{e} = \mathbf{b}
\label{eq:maxwells_equations_A}.
\end{equation}
Here, $\hat{\epsilon}_r$ describes the spatial distribution of the relative permittivity ($\epsilon_r$), $k_0=\omega^2/c^2$ is the free-space wavenumber, $\mathbf{e}$ is the electric field distribution, $\mathbf{j}$ is the electric current density, and $\hat{A} = \hat{A}^T$ due to Lorentz reciprocity. Eq. (\ref{eq:maxwells_equations_A}) is the starting point of the finite-difference frequency-domain (FDFD) simulation technique \cite{shin2012choice}, where it is discretized on a spatial grid, and the electric field $\mathbf{e}$ is solved given a particular permittivity distribution, $\boldsymbol{\epsilon}_r$, and source, $\mathbf{b}$.

To relate this formulation to the transfer matrix $\hat{W}$, we now define source terms $\mathbf{b}_i$, $i \in 1 \dots 2N$, that correspond to a source placed in one of the input or output ports.  Here we assume a total of $N$  input and $N$ output waveguides.  The spatial distribution of the source term, $\mathbf{b}_i$, matches the mode of the $i$-th single-mode waveguide. Thus, the electric field amplitude in port $i$ is given by $\mathbf{b}_i^{\ T} \mathbf{e}$, and we may establish a relationship between $\mathbf{e}$ and $\mathbf{X}_\textrm{in}$, as
\begin{equation}
X_{\textrm{in},i} = \mathbf{b}_i^{\ T} \mathbf{e}
\end{equation}
for $i = 1~...~N$ over the input port indices, where $X_{\textrm{in},i}$ is the $i$-th component of $\mathbf{X}_\textrm{in}$. Or more compactly, 
\begin{equation}
\mathbf{X}_\textrm{in} \equiv \hat{P}_{\textrm{in}} \mathbf{e},
\label{eq:basis_conversion_in}
\end{equation}
Similarly, we can define 
\begin{equation}
Z_{\textrm{out},i} = \mathbf{b}_{i+N}^{\ T} \mathbf{e}
\end{equation}
for $i+N = (N+1)~...~2N$ over the output port indices, or, 
\begin{equation}
\mathbf{Z}_\textrm{out} \equiv \hat{P}_{\textrm{out}} \mathbf{e},
\label{eq:basis_conversion_in}
\end{equation}
and, with this notation, Eq. (\ref{eq:original_linear_system}) becomes
\begin{equation}
\hat{W}\hat{P}_{\textrm{in}}\mathbf{e} = \hat{P}_{\textrm{out}}\mathbf{e}
\label{eq:summary_of_basis_change}
\end{equation}


%

We now use the above definitions to evaluate the cost function gradient in Eq. (\ref{eq:backprop_general}). In particular, with Eqs. (\ref{eq:backprop_general}) and (\ref{eq:summary_of_basis_change}), we arrive at 
\begin{equation}
\frac{d\mathcal{L}}{d\epsilon_l} = -\mathcal{R}\left\{\boldsymbol{\delta_l}^T \hat{P}_{\textrm{out}} \hat{A}^{-1} \frac{d\hat{A}}{d\epsilon_l} \hat{A}^{-1} \mathbf{b}_{x,l-1} \right\}.
\label{eq:objective_function_derivative_EM_basis}
\end{equation}
Here $\mathbf{b}_{x,l-1}$ is the modal source profile that creates the input field amplitudes $\mathbf{X}_{l-1}$ at the input ports. 

The key insight of the adjoint variable method is that we may interpret this expression as an operation involving the field solutions of two electromagnetic simulations, which we refer to as the `original' (og) and the `adjoint' (aj)
\begin{align}
\hat{A} \mathbf{e}_{\textrm{og}} &= \mathbf{b}_{x,l-1} \\
\hat{A} \mathbf{e}_{\textrm{aj}} &= \hat{P}_{\textrm{out}}^T \boldsymbol{\delta},
\label{eq:adjoint}
\end{align}
where we have made use of the symmetric property of $\hat{A}$. 

Eq. (\ref{eq:objective_function_derivative_EM_basis}) can now be expressed in a compact form as 
\begin{equation}
\frac{d\mathcal{L}}{d\epsilon_l} = -\mathcal{R}\left\{\mathbf{e}_{\textrm{aj}}^T\frac{d\hat{A}}{d\epsilon_l}\mathbf{e}_{\textrm{og}} \right\}.
\label{eq:objective_function_derivative_simple}
\end{equation}

If we assume that this phase shifter spans a set of points, $\mathbf{r}_\phi$ in our system, then, from Eq. (\ref{eq:maxwells_equations_physics}), we obtain
\begin{equation}
\frac{d\hat{A}}{d\epsilon_l} = -k_0^2 \sum_{{\mathbf{r}\,}' \in \mathbf{r}_\phi}\hat{\delta}_{\mathbf{r},{\mathbf{r}\,}'},
\label{eq:dAdphi}
\end{equation}
where $\hat{\delta}_{\mathbf{r},{\mathbf{r}\,}'}$ is the Kronecker delta.

Inserting this into Eq. (\ref{eq:objective_function_derivative_simple}), we thus find that the gradient is given by the overlap of the two fields over the phase-shifter positions
\begin{equation}
\frac{d\mathcal{L}}{d\epsilon_l} = k_0^2 \mathcal{R}\left\{ \sum_{\mathbf{r} \in \mathbf{r}_\phi} \mathbf{e}_{\textrm{aj}}(\mathbf{r}) \mathbf{e}_{\textrm{og}}(\mathbf{r}) \right\}.
\label{eq:sensitivity_sum}
\end{equation}
This result now allows for the computation in parallel of the gradient of the loss function with respect to \textit{all} phase shifters in the system, given knowledge of the original and adjoint fields.

\begin{figure}[t]
\includegraphics[width=0.9\columnwidth]{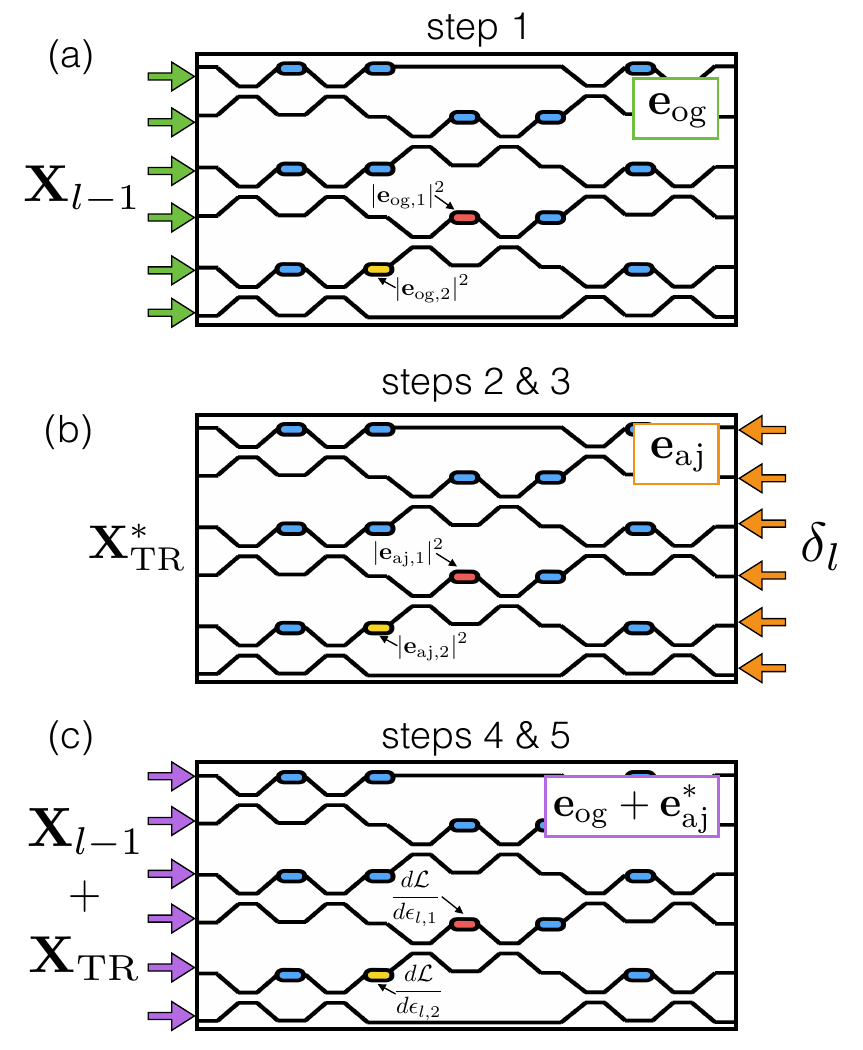}
\caption{\label{fig:TR_schematic} Schematic illustration of our proposed method for experimental measurement of gradient information. The box region represents the OIU.  The colored ovals represent tunable phase shifters, and we illustrate computing the gradient with respect to the red and the yellow phase shifters, labeled $1$ and $2$, respectively. (a): We send the original set of amplitudes $\mathbf{X}_{l-1}$ and measure the constant intensity terms at each phase shifter. (b): We send the adjoint mode amplitudes, given by $\boldsymbol{\delta}_l$, through the output side of our device, recording $\mathbf{X}_{TR}^*$ from the opposite side, as well as $|\mathbf{e}_\textrm{aj}|^2$ in each phase-shifter. (c): We send in $\mathbf{X}_{l-1}$ + $\mathbf{X}_{TR}$, interfering $\mathbf{e}_{\textrm{og}}$ and $\mathbf{e}_\textrm{aj}^*$ inside the device and recovering the gradient information for all phase shifters simultaneously.}
\end{figure}

\section{Experimental measurement of gradient} \label{sec:experimental}

We now introduce our time-reversal interference method (TRIM) for computing the gradient from the previous section through \textit{in situ} intensity measurements.  This represents the most significant result of this paper.  Specifically, we wish to generate an intensity pattern with the form $\mathcal{R}\big\{\mathbf{e}_\textrm{og} \mathbf{e}_\textrm{aj} \big\}$, matching that of Eq. (\ref{eq:sensitivity_sum}).  We note that interfering $\mathbf{e}_{\textrm{og}}$ and $\mathbf{e}_\textrm{aj}^{\,*}$ directly in the system results in the intensity pattern:
\begin{equation}
I = |\mathbf{e}_\textrm{og}|^2 + |\mathbf{e}_\textrm{aj}|^2 + 2\mathcal{R}\big\{\mathbf{e}_\textrm{og}\mathbf{e}_{\textrm{aj}} \big\},
\label{eq:intensity_correct}
\end{equation}
the last term of which matches Eq. (\ref{eq:sensitivity_sum}). Thus, the gradient can be computed purely through intensity measurements if the field $\mathbf{e}_\textrm{aj}^{\,*}$ can be generated in the OIU.

\begin{figure*}[t]
\includegraphics[width=\textwidth]{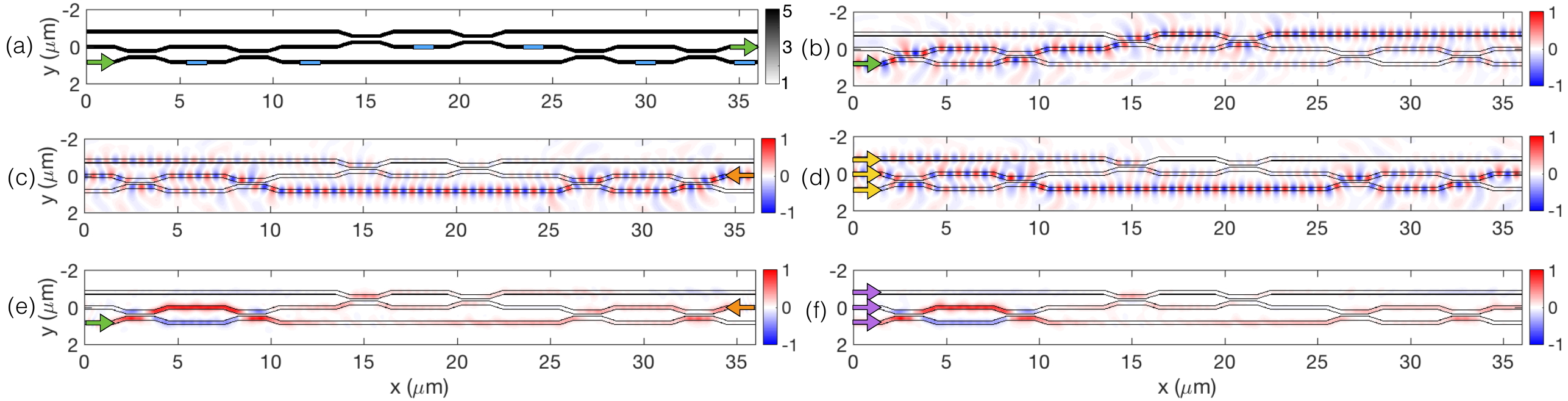}
\caption{\label{fig:fdfd} Numerical demonstration of the TRIM method of Section \ref{sec:experimental}.  (a): Relative permittivity distribution for three MZIs arranged to perform a 3x3 linear operation. Blue boxes represent where phase shifters would be placed in this system. As an example, we compute the gradient information for a layer with $\mathbf{X}_{l-1} = [0~0~1]^T$ and $\boldsymbol{\delta}_l = [0~1~0]^T$, corresponding to the bottom left and middle right port, respectively. (b): Real part of the simulated electric field $E_z$ corresponding to injection from the bottom left port. (c): Real part of the adjoint $E_z$, corresponding to injection from the middle right port. (d): Time-reversed adjoint field as constructed by our method, fed in through all three ports on the left. (e): The gradient information $\frac{d\mathcal{L}}{d\epsilon_l}(x,y)$ as obtained directly by the adjoint method, normalized by its maximum absolute value. (f): The gradient information as obtained by the method introduced in this work, normalized by its maximum absolute value. Namely, the field pattern from (b) is interfered with the time-reversed adjoint field of (d) and the constant intensity terms are subtracted from the resulting intensity pattern. Panels (e) and (f) match with high precision.}
\end{figure*}

The adjoint field for our problem, $\mathbf{e}_\textrm{aj}$, as defined in Eq. (\ref{eq:adjoint}), is sourced by $\hat{P}_{\textrm{out}}^T \boldsymbol{\delta}_l$, meaning that it physically corresponds to a mode sent into the system from the output ports.  As complex conjugation in the frequency domain corresponds to time-reversal of the fields, we expect $\mathbf{e}^{\,*}_{\textrm{aj}}$ to be sent in from the input ports. Formally, to generate $\mathbf{e}^{\,*}_{\textrm{aj}}$, we wish to find a set of input source amplitudes, $\mathbf{X}_{TR}$, such that the output port source amplitudes, $\mathbf{Z}_{TR} = \hat{W}\mathbf{X}_{TR}$, are equal to the complex conjugate of the adjoint amplitudes, or $\boldsymbol{\delta}_l^*$. Using the unitarity property of transfer matrix $\hat{W}_l$ for a lossless system, along with the fact that $\hat{P}_{\textrm{out}} \hat{P}_{\textrm{out}}^T = \hat{I}$ for output modes, the input mode amplitudes for the time-reversed adjoint can be computed as
\begin{equation}
\mathbf{X}_{TR}^* = \hat{W}_l^T\boldsymbol{\delta}_l.
\label{eq:TR_adjoint}
\end{equation}
As discussed earlier, $\hat{W}_l^T$ is the transfer matrix from output ports to input ports. Thus, we can experimentally determine $\mathbf{X}_{TR}$ by sending $\boldsymbol{\delta}_l$ into the device output ports, measuring the output at the input ports, and taking the complex conjugate of the result.

We now summarize the procedure for experimentally measuring the gradient of an OIU layer in the ANN with respect to the permittivities of this layer's integrated phase shifters:
\begin{enumerate}
  \itemsep0em 
  \item Send in the original field amplitudes $\mathbf{X}_{l-1}$ and measure and store the intensities at each phase shifter.
  \item Send $\boldsymbol{\delta}_l$ into the output ports and measure and store the intensities at each phase shifter.
  \item Compute the time-reversed adjoint input field amplitudes as in Eq. (\ref{eq:TR_adjoint}).
  \item Interfere the original and the time-reversed adjoint fields in the device, measuring again the resulting intensities at each phase shifter.
  \item Subtract the constant intensity terms from steps 1 and 2 and multiply by $k_0^2$ to recover the gradient as in Eq. (\ref{eq:sensitivity_sum}).
\end{enumerate}



This procedure is also illustrated in Fig. \ref{fig:TR_schematic}.

\section{Numerical Gradient Demonstration} \label{sec:numerical}

We numerically demonstrate this procedure in Fig. \ref{fig:fdfd} with a series of FDFD simulations of an OIU implementing a $3 \times 3$ unitary matrix \cite{Reck1994}. These simulations are intended to represent the gradient computation corresponding to one OIU in a single layer, $l$, of a neural network with input $\mathbf{X}_{l-1}$ and delta vector $\boldsymbol{\delta}_l$. In these simulations, we use absorbing boundary conditions on the outer edges of the system to eliminate back-reflections.  The relative permittivity distribution is shown in Fig. \ref{fig:fdfd}(a) with the positions of the variable phase shifters in blue.  For demonstration, we simulate a specific case where $\mathbf{X}_{l-1} = [0~0~1]^T$, with unit amplitude in the bottom port and we choose $\boldsymbol{\delta}_l = [0~1~0]^T$.  In Fig. \ref{fig:fdfd}(b), we display the real part of $\textbf{e}_\textrm{og}$, corresponding to the original, forward field.  

The real part of the adjoint field, $\textbf{e}_\textrm{aj}$, corresponding to the cost function $\mathcal{L} = \mathcal{R}\left\{\boldsymbol{\delta}_l^T \hat{W}_l \mathbf{X}_{l-1} \right\}$ is shown in Fig. \ref{fig:fdfd}(c).  In Fig. \ref{fig:fdfd}(d) we show the real part of the time-reversed copy of $\textbf{e}_\textrm{aj}$ as computed by the method described in the previous section, in which $\mathbf{X}^*_{TR}$ is sent in through the input ports.  There is excellent agreement, up to a constant, between the complex conjugate of the field pattern of (c) and the field pattern of (d), as expected.  

In Fig. \ref{fig:fdfd}(e), we display the gradient of the objective function with respect to the permittivity of each point of space in the system, as computed with the adjoint method, described in Eq. (\ref{eq:sensitivity_sum}).  In Fig. \ref{fig:fdfd}(f), we show the same gradient information, but instead computed with the method described in the previous section.  Namely, we interfere the field pattern from panel (b) with the field pattern from panel (d), subtract constant intensity terms, and multiply by the appropriate constants.  Again, (b) and (d) agree with good precision.

We note that in a realistic system, the gradient must be constant for any stretch of waveguide between waveguide couplers because the interfering fields are at the same frequency and are traveling in the same direction.  Thus, there should be no distance dependence in the corresponding intensity distribution. This is largely observed in our simulation, although small fluctuations are visible because of the proximity of the waveguides and the sharp bends, which were needed to make the structure compact enough for simulation within a reasonable time. In practice, the importance of this constant intensity is that it can be detected \textit{after} each phase shifter, instead of inside of it.

Finally, we note that this numerically generated system experiences a total power loss of 41\% due to scattering caused by very sharp bends and stair-casing of the structure in the simulation.  We also observe approximately 5-10\% mode-dependent loss, as determined by measuring the difference in total transmitted power corresponding to injection at different input ports.  Minimal amounts of reflection are also visible in the field plots.  Nevertheless, TRIM still reconstructs the adjoint sensitivity with very good fidelity.


\section{Example of ANN training} \label{sec:ANN_simulation}

\begin{figure}
\includegraphics[width=\columnwidth]{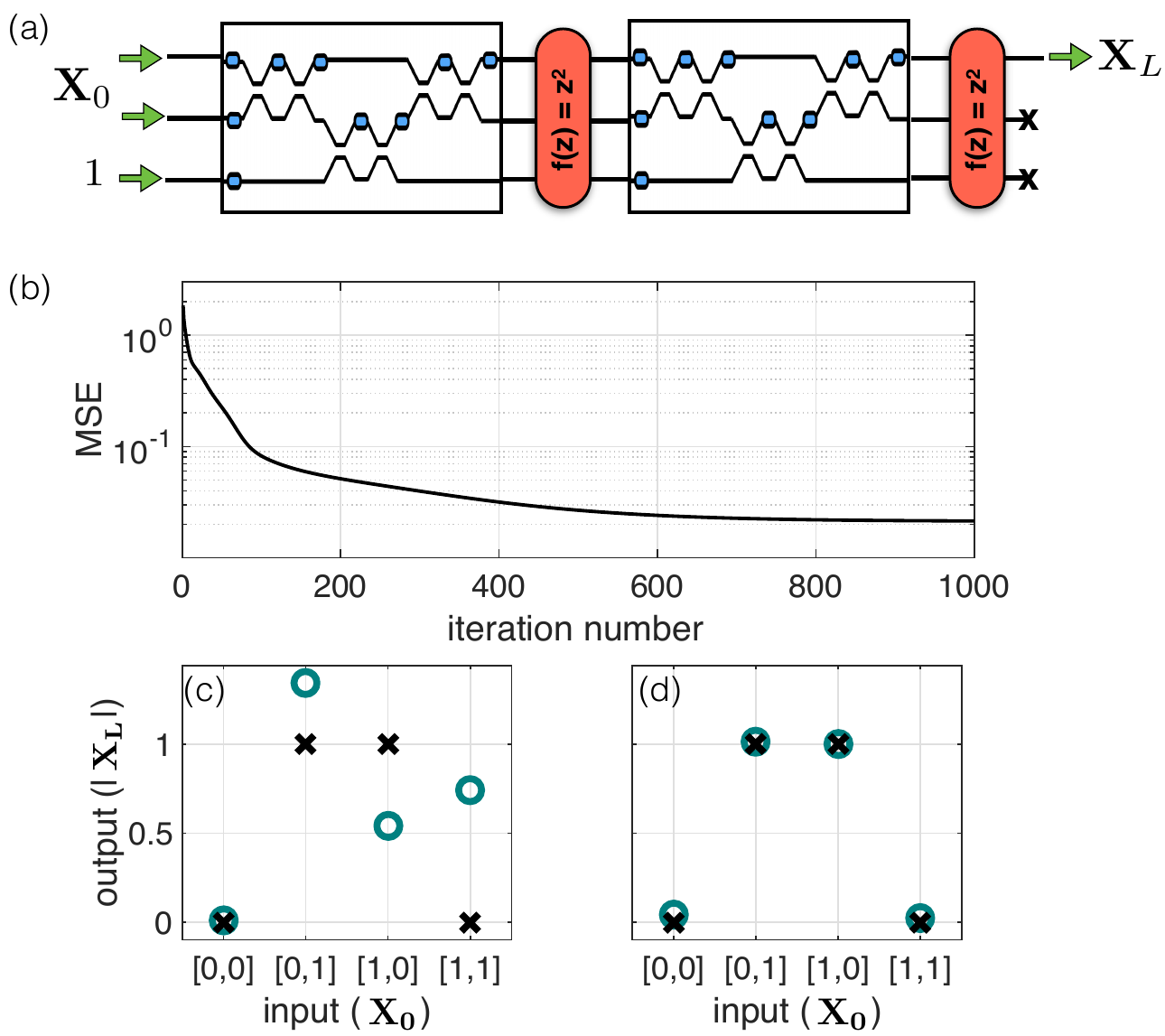}
\caption{Numerical demonstration of a photonic ANN implementing an XOR gate using the backpropagation algorithm and adjoint method described in this work. (a) The architecture of the ANN.  Two layers of $3 \times 3$ OIUs with $z^2$ activations.  (b) The mean-squared error (MSE) between the predictions and targets as a function of training iterations. (c) The absolute value of the network predictions (blue circles) and targets (black crosses) before training.  (d) The absolute value of the network predictions after training, showing that the network has successfully learned the XOR function. \label{fig:demo} }
\end{figure}


Finally, we use the techniques from the previous Sections to numerically demonstrate the training of a photonic ANN to implement a logical XOR gate, defined by the following input to target ($\mathbf{X}_0 \to \mathbf{T}$) pairs
\begin{equation}
[0~0]^T \to 0,  ~~~ [0~1]^T \to 1,  ~~~ [1~0]^T \to 1, ~~~ [1~1]^T \to 0.
\label{eq:XOR_definition}
\end{equation}
This problem was chosen as demonstration of learning a nonlinear mapping from input to output \cite{vandoorne2014experimental} and is simple enough to be solved with a small network with only four training examples.

As diagrammed in Fig. \ref{fig:demo}a, we choose a network architecture consisting of two $3 \times 3$ unitary OIUs.  On the forward propagation step, the binary representation of the inputs, $\textbf{X}_0$, is sent into the first two input elements of the ANN and a constant value of $1$ is sent into the third input element, which serves to introduce artificial bias terms into the network.  These inputs are sent through a $3 \times 3$ unitary OIU and then the element-wise activation $f(z) = z^2$ is applied.  The output of this step is sent to another $3 \times 3$ OIU and sent through another activation of the same form.  Finally, the first output element is taken to be our prediction, $\textbf{X}_L$, ignoring the last two output elements.  Our network is repeatedly trained on the four training examples defined in Eq. (\ref{eq:XOR_definition}) and using the mean-squared cost function presented in Eq. (\ref{eq:NN_forward}).

For this demonstration, we utilized a matrix model of the system, as described in \cite{Reck1994,Clements2016}, with mathematical details described in Appendix B.  This model allows us to compute an output of the system given an input mode and the settings of each phase shifter.  Although this is not a first-principle electromagnetic simulation of the system, it provides information about the complex fields at specific reference points within the circuit, which enables us to implement training using the backpropagation method as described in Section \ref{sec:ANN}, combined with the adjoint gradient calculation from Section \ref{sec:adjoint}.  Using these methods, at each iteration of training we compute the gradient of  our cost function with respect to the phases of each of the integrated phase shifters, and sum them over the four training examples.  Then, we perform a simple steepest-descent update to the phase shifters, in accordance with the gradient information. This is consistent with the standard training protocol for an ANN implemented on a conventional computer.  Our network successfully learned the XOR gate in around 400 iterations. The results of the training are shown in Fig. \ref{fig:demo}b-d.

We note that this is meant to serve as a simple demonstration of using the \textit{in-situ} backpropagation technique for computing the gradients needed to train photonic ANNs.  However, our method may equally be performed on more complicated tasks, which we show in the Appendix C.

\section{Discussion and Conclusion} \label{sec:conclusion}


 
Here, we justify some of the assumptions made in this work.  Our strategy for training a photonic ANN relies on the ability to create arbitrary complex inputs.  We note that a device for accomplishing this has been proposed and discussed in \cite{Miller2017}.  Our recovery technique further requires an integrated intensity detection scheme to occur in parallel and with virtually no loss.  This may be implemented by integrated, transparent photo-detectors, which have already been demonstrated in similar systems \cite{Annoni2017}.  Furthermore, as discussed, this measurement may occur in the waveguide regions directly after the phase shifters, which eliminates the need for phase shifter and photodetector components at the same location.  Finally, in our procedure for experimentally measuring the gradient information, we suggested running isolated forward and adjoint steps, storing the intensities at each phase shifter for each step, and then subtracting this information from the final interference intensity. Alternatively, one may bypass the need to store these constant intensities by introducing a low-frequency modulation on top of one of the two interfering fields in Fig. \ref{fig:TR_schematic}(c), such that the product term of Eq. (\ref{eq:intensity_correct}) can be directly measured from the low-frequency signal.  A similar technique was used in \cite{Annoni2017}.


We now discuss some of the limitations of our method.  In the derivation, we had assumed the $\hat{W}$ operator to be unitary, which corresponds to a lossless OIU.  In fact, we note that our procedure is exact in the limit of a lossless, feed-forward, and reciprocal system.  However, with the addition of any amount of uniform loss, $\hat{W}$ is still unitary up to a constant, and our procedure may still be performed with the added step of scaling the measured gradients depending on this loss (see a related discussion in Ref. \cite{Miller2017}).  Uniform loss conditions are satisfied in the OIUs experimentally demonstrated in Refs. \cite{Shen2017, Miller2013}.  Mode-dependent loss, such as asymmetry in the MZI mesh layout or fabrication errors, should be avoided as its presence limits the ability to accurately reconstruct the time-reversed adjoint field.  Nevertheless, our simulation in Fig. \ref{fig:fdfd} indicates that an accurate gradient can be obtained even in the presence of significant mode-dependent loss.  In the experimental structures of  Refs. \cite{Shen2017, Miller2013}, the mode-dependent loss is made much lower due to the choice of the MZI mesh.  Thus we expect our protocol to work in practical systems.  Our method, in principle, computes gradients in parallel and scales in constant time.  In practice, to get this scaling would require careful design of the circuits controlling the OIUs.

Conveniently, since our method does not directly assume any specific model for the linear operations, it may gracefully handle imperfections in the OIUs, such as deviations from perfect 50-50 splits in the MZIs.  Lastly, while we chose to make an explicit distinction between the input ports and the output ports, i.e. we assume no backscattering in the system, this requirement is not strictly necessary. Our formalism can be extended to the full scattering matrix.  However, this would require special treatment for subtracting the backscattering. 

The problem of overfitting is one that must be addressed by `regularization' in any practical realization of a neural network.  Photonic ANNs of this class provide a convenient approach to regularization based on `dropout' \cite{srivastava2014dropout}.  In the dropout procedure, certain nodes are probabilistically and temporarily `deleted’ from the network during train time, which has the effect of forcing the network to find alternative paths to solve the problem at hand.  This has a strong regularization effect and has become popular in conventional ANNs.  Dropout may be implemented simply in the photonic ANN by `shutting off’ channels in the activation functions during training.  Specifically, at each time step and for each layer $l$ and element $i$, one may set $f_l(Z_i) = 0$ with some fixed probability.


In conclusion, we have demonstrated a method for performing backpropagation in an ANN based on a photonic circuit.  This method works by physically propagating the adjoint field and interfering its time-reversed copy with the original field.  The gradient information can then be directly measured out as an \textit{in-situ} intensity measurement. While we chose to demonstrate this procedure in the context of ANNs, it is broadly applicable to any reconfigurable photonic system.  One could imagine this setup being used to tune phased arrays \cite{sun2013large}, optical delivery systems for dielectric laser accelerators \cite{hughes2017chip}, or other systems that rely on large meshes of integrated optical phase shifters.  Furthermore, it may be applied to sensitivity analysis of photonic devices, enabling spatial sensitivity information to be measured as an intensity in the device.

Our work should enhance the appeal of photonic circuits in deep learning applications, allowing for training to happen directly inside the device in an efficient and scalable manner.  Furthermore, this method is broadly applicable to integrated and adaptive optical systems, enabling the possibility for automatic self-configuration and optimization without resorting to brute force gradient computation or model-based methods, which often do not perfectly represent the physical system.

\section*{Funding Information}
Gordon and Betty Moore Foundation (GBMF4744); Schweizerischer Nationalfonds zur Förderung der Wissenschaftlichen Forschung (P300P2\_177721);  Air Force Office of Scientific Research (FA9550-17-1-0002).



\bibliography{photonic_nn}
\pagebreak
\appendix

\section{Non-holomorphic Backpropagation}
\label{noholo}

In the previous derivation, we have assumed that the functions $\mathbf{f}_l(\cdot)$ are holomorphic.  For each element of input $\mathbf{Z}_l$, labeled $z$, this means that the derivative of $\mathbf{f}_l(z)$ with respect to its complex argument is well defined, or the derivative 
\begin{equation}
\frac{d\mathbf{f}_l}{dz} = \lim_{\Delta z \to 0} \frac{\mathbf{f}_l(z+\Delta z) - \mathbf{f}_l(z-\Delta z)}{2 \Delta z}
\end{equation}
does not depend on the direction that $\Delta z$ approaches $0$ in the complex plane.  

Here we show how to extend the backpropagation derivation to non-holomorphic activation functions.  We first examine the starting point of the backpropagation algorithm, considering the change in the mean-squared loss function with respect to the permittivity of a phase shifter in the last layer OIU, as written in Eq. (7) of the main text as 
\begin{equation}
\frac{d\mathcal{L}}{d\epsilon_L} = \mathcal{R}\left\{\boldsymbol{\Gamma}_L^T \frac{d\mathbf{X}_L}{d\epsilon_L} \right\}
\label{eq:beginning}
\end{equation}
Where we had defined the error vector $\boldsymbol{\Gamma}_L \equiv \big(\mathbf{X}_L - \mathbf{T} \big)^*$ for simplicity and $\mathbf{X}_L = \mathbf{f}_L(\hat{W}_L \mathbf{X}_{L-1})$ is the output of the final layer.

To evaluate this expression for non-holomorphic activation functions, we split $\mathbf{f}_L(\mathbf{Z})$ and its argument into their real and imaginary parts
\begin{equation}
\mathbf{f}_L(\mathbf{Z}) = \mathbf{u}(\boldsymbol{\alpha}, \boldsymbol{\beta}) + i \mathbf{v}(\boldsymbol{\alpha}, \boldsymbol{\beta}),
\end{equation}
where $i$ is the imaginary unit and $\boldsymbol{\alpha}$ and $\boldsymbol{\beta}$ are the real and imaginary parts of $\mathbf{Z}_L$, respectively.

We now wish to evaluate $\frac{d\mathbf{X}_L}{d\epsilon_L} = \frac{d\mathbf{f}_L(\mathbf{Z})}{d\epsilon_L}$, which gives the following via the chain rule

\begin{equation}
\frac{d\mathbf{f}}{d\epsilon} = \frac{d\mathbf{u}}{d\boldsymbol{\alpha}} \odot \frac{d\boldsymbol{\alpha}}{d\epsilon} + \frac{d\mathbf{u}}{d\boldsymbol{\beta}} \odot \frac{d\boldsymbol{\beta}}{d\epsilon} + i\frac{d\mathbf{v}}{d\boldsymbol{\alpha}} \odot \frac{d\boldsymbol{\alpha}}{d\epsilon} + i\frac{d\mathbf{v}}{d\boldsymbol{\beta}} \odot \frac{d\boldsymbol{\beta}}{d\epsilon},
\end{equation}
where we have dropped the layer index for simplicity.  Here, terms of the form $\frac{d\mathbf{x}}{d\mathbf{y}}$ correspond to element-wise differentiation of the vector $\mathbf{x}$ with respect to the vector $\mathbf{y}$.  For example, the $i$-th element of the vector $\frac{d\mathbf{x}}{d\mathbf{y}}$ is given by $\frac{dx_i}{dy_i}$.

Now, inserting into Eq. (\ref{eq:beginning}), we have 
\begin{align}
\frac{d\mathcal{L}}{d\epsilon_L} = \mathcal{R}\Bigg\{  &\boldsymbol{\Gamma}_L^T \odot \left( \frac{d\mathbf{u}}{d\boldsymbol{\alpha}} + i  \frac{d\mathbf{v}}{d\boldsymbol{\alpha}} \right)^T \frac{d\boldsymbol{\alpha}}{d\epsilon_L} \\
+ &\boldsymbol{\Gamma}_L^T \odot \left( \frac{d\mathbf{u}}{d\boldsymbol{\beta}} + i  \frac{d\mathbf{v}}{d\boldsymbol{\beta}} \right)^T \frac{d\boldsymbol{\beta}}{d\epsilon_L}  \Bigg\}.
\end{align}

We now define real and imaginary parts of $\boldsymbol{\Gamma}_L$ as $\boldsymbol{\Gamma}_\textrm{R}$ and $\boldsymbol{\Gamma}_\textrm{I}$, respectively.  Inserting the definitions of $\boldsymbol{\alpha}$ and $\boldsymbol{\beta}$ in terms of $\hat{W}_L$ and $\mathbf{X}_{L-1}$ and doing some algebra, we recover
\begin{align}
\frac{d\mathcal{L}}{d\epsilon_L} = \mathcal{R}\Bigg\{ &\left( \boldsymbol{\Gamma}_\textrm{R} \odot \frac{d\mathbf{u}}{d\boldsymbol{\alpha}} \right)^T \frac{d\hat{W}_L}{d\epsilon_L} \mathbf{X}_{L-1} \\
- &\left( \boldsymbol{\Gamma}_\textrm{I} \odot \frac{d\mathbf{v}}{d\boldsymbol{\alpha}} \right)^T \frac{d\hat{W}_L}{d\epsilon_L} \mathbf{X}_{L-1} \\
-i &\left( \boldsymbol{\Gamma}_\textrm{R} \odot \frac{d\mathbf{u}}{d\boldsymbol{\beta}} \right)^T \frac{d\hat{W}_L}{d\epsilon_L} \mathbf{X}_{L-1} \\
+i &\left( \boldsymbol{\Gamma}_\textrm{I} \odot \frac{d\mathbf{v}}{d\boldsymbol{\beta}} \right)^T \frac{d\hat{W}_L}{d\epsilon_L} \mathbf{X}_{L-1}
\Bigg\}.
\label{eq:final_answer}
\end{align}

Finally, the expression simplifies to 
\begin{align}
\frac{d\mathcal{L}}{d\epsilon_L} = \mathcal{R}\Bigg\{ \Bigg[ & \boldsymbol{\Gamma}_\textrm{R} \odot \left( \frac{d\mathbf{u}}{d\boldsymbol{\alpha}} - i\frac{d\mathbf{u}}{d\boldsymbol{\beta}} \right) \\
 + & \boldsymbol{\Gamma}_\textrm{I} \odot \left( -\frac{d\mathbf{v}}{d\boldsymbol{\alpha}} + i\frac{d\mathbf{v}}{d\boldsymbol{\beta}} \right)  \Bigg]^T \frac{d\hat{W}_L}{d\epsilon_L} \mathbf{X}_{L-1} \Bigg\}.
\end{align}

As a check, if we insert the conditions for $\mathbf{f}_L(\mathbf{Z})$ to be holomorphic, namely
\begin{equation}
\frac{d\mathbf{u}}{d\boldsymbol{\alpha}} = \frac{d\mathbf{v}}{d\boldsymbol{\beta}}, ~~~\textrm{and}~~~ \frac{d\mathbf{u}}{d\boldsymbol{\beta}} = -\frac{d\mathbf{v}}{d\boldsymbol{\alpha}},
\end{equation}
Eq. (\ref{eq:final_answer}) simplifies to 
\begin{align}
\frac{d\mathcal{L}}{d\epsilon_L}
&= \mathcal{R}\Bigg\{ \Bigg[ \boldsymbol{\Gamma}_\textrm{R} \odot \left( \frac{d\mathbf{u}}{d\boldsymbol{\alpha}} + i\frac{d\mathbf{v}}{d\boldsymbol{\alpha}} \right) + \\
&\boldsymbol{\Gamma}_\textrm{I} \odot \left( -\frac{d\mathbf{v}}{d\boldsymbol{\alpha}} + i\frac{d\mathbf{u}}{d\boldsymbol{\alpha}} \right) \Bigg]^T \frac{d\hat{W}_L}{d\epsilon_L} \mathbf{X}_{L-1}
\Bigg\} \\
&= \mathcal{R}\Bigg\{ \left[ \boldsymbol{\Gamma}_L \odot \left( \frac{d\mathbf{u}}{d\boldsymbol{\alpha}} + i\frac{d\mathbf{v}}{d\boldsymbol{\alpha}} \right) \right]^T \frac{d\hat{W}_L}{d\epsilon_L} \mathbf{X}_{L-1} \Bigg\}\\
&= \mathcal{R}\Bigg\{ \left[ \boldsymbol{\Gamma}_L \odot {\mathbf{f}_l}^{'}(\mathbf{Z}_{L}) \right]^T \frac{d\hat{W}_L}{d\epsilon_L} \mathbf{X}_{L-1}
\Bigg\}\\
&= \mathcal{R}\Bigg\{ \bm{\delta}_L^T \frac{d\hat{W}_L}{d\epsilon_L} \mathbf{X}_{L-1}
\Bigg\}
\end{align}
as before.

This derivation may be similarly extended to any layer $l$ in the network.  For holomorphic activation functions, whereas we originally defined the $\boldsymbol{\delta}$ vectors as 
\begin{equation}
\boldsymbol{\delta}_l = \boldsymbol{\Gamma}_{l} \odot {\mathbf{f}_l}^{'}(\mathbf{Z}_{l}),
\label{eq:delta_def_gen}
\end{equation}
for non-holomorphic activation functions, the respective definition is 
\begin{equation}
\boldsymbol{\delta}_l = \boldsymbol{\Gamma}_R \odot \left( \frac{d\mathbf{u}}{d\boldsymbol{\alpha}} - i\frac{d\mathbf{u}}{d\boldsymbol{\beta}} \right)
 + \boldsymbol{\Gamma}_\textrm{I} \odot \left( -\frac{d\mathbf{v}}{d\boldsymbol{\alpha}} + i\frac{d\mathbf{v}}{d\boldsymbol{\beta}} \right),
\end{equation}
where $\boldsymbol{\Gamma}_\textrm{R}$ and $\boldsymbol{\Gamma}_\textrm{I}$ are the respective real and imaginary parts of $\bm{\Gamma}_l$, $\mathbf{u}$ and $\mathbf{v}$ are the real and imaginary parts of $\mathbf{f}_l(\cdot)$, and $\boldsymbol{\alpha}$ and $\boldsymbol{\beta}$ are the real and imaginary parts of $\mathbf{Z}_l$, respectively.

We can write this more simply as 
\begin{equation}
\boldsymbol{\delta}_l = \mathcal{R}\left\{\bm{\Gamma}_l \odot \frac{d\mathbf{f}}{d\bm{\alpha}} \right\} - i \mathcal{R}\left\{ \bm{\Gamma}_l \odot \frac{d\mathbf{f}}{d\bm{\beta}}\right\}.
\end{equation}

In polar coordinates where $\mathbf{Z} = \mathbf{r}\exp{(i\bm{\phi})}$ and $\mathbf{f} = \mathbf{f}(\mathbf{r},\bm{\phi})$, this equation becomes
\begin{equation}
\boldsymbol{\delta}_l = \exp{(-i\bm{\phi})}\left(\mathcal{R}\left\{\bm{\Gamma}_l \odot \frac{d\mathbf{f}}{d\mathbf{r}} \right\} - i\mathcal{R}\left\{ \bm{\Gamma}_l \odot \frac{1}{\mathbf{r}}\frac{d\mathbf{f}}{d\bm{\phi}} \right\}  \right) \label{eqn:noholo}
\end{equation}
where all operations are element-wise.

\section{Photonic neural network simulation}
\label{pnn}

\begin{figure}[t]
\centering
\includegraphics[width=\columnwidth, , trim = 0in 0in 3.6in 1.2in, clip = true]{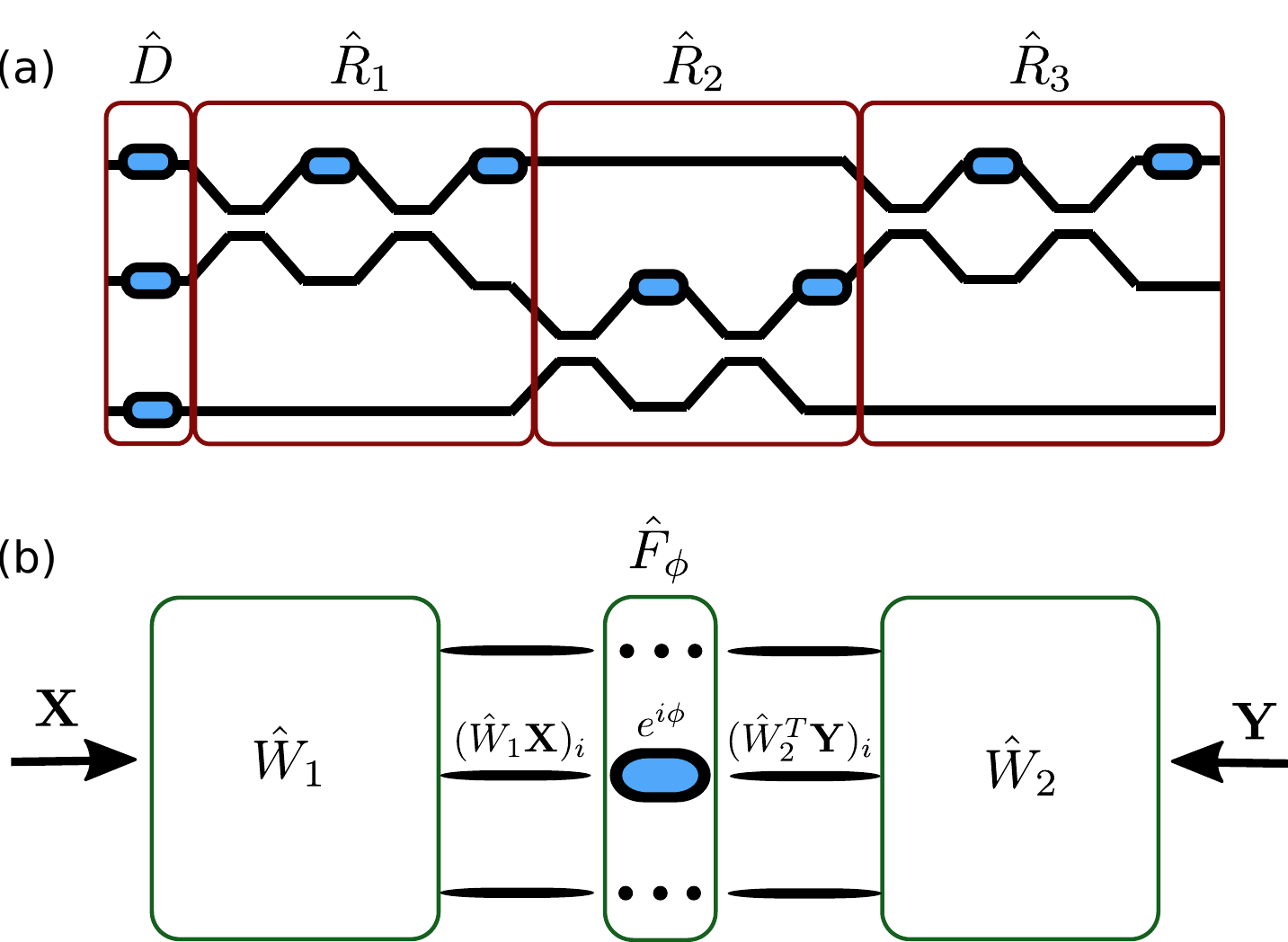}
\caption{\label{fig:W_opt}(a): An OIU implementing a universal $3\times 3$ unitary operation, parameterized as in Ref. \cite{Clements2016}. (b): Illustration of Eq. (\ref{eqn:XY}) for the computation of the gradient with respect to a given phase shifter.}
\end{figure}

In Sections 4 and 5 of the main text, we have shown, starting from Maxwell's equations, how the gradient information defined for an arbitrary problem can be obtained through electric field intensity measurements. However, since the full electromagnetic problem is too large to solve repeatedly, for the purposes of demonstration of a functioning neural network, in Section 6 we use the analytic, matrix representation of a mesh of MZIs as described in Ref. \cite{Clements2016}. Namely, for an even $N$, the matrix $\hat{W}$ of the OIU is parametrized as the product of $N + 1$ unitary matrices:
\begin{equation}
\hat{W} = \hat{R}_N \hat{R}_{N-1} \dots \hat{R}_2 \hat{R}_1 \hat{D} ,
\label{eq:W_R}
\end{equation}
where each $\hat{R}_i$ implements a number of two-by-two unitary operations corresponding to a given MZI, and $\hat{D}$ is a diagonal matrix corresponding to an arbitrary phase delay added to each port. This is schematically illustrated in Fig. \ref{fig:W_opt}(a) for $N = 3$. For the ANN training, we need to compute terms of the form 
\begin{equation}
\frac{d\mathcal{L}}{d\phi} =
\mathcal{R}\left\{ \mathbf{Y}^T \frac{d \hat{W}}{d\phi} \mathbf{X} \right\},
\label{eqn:L}
\end{equation}
for an arbitrary phase $\phi$ and vectors $\mathbf{X}$ and $\mathbf{Y}$ defined following the steps in the main text. Because of the feed-forward nature of the OIU-s, the matrix $\hat{W}$ can also be split as
\begin{equation}
\hat{W} = \hat{W}_2 \hat{F_\phi} \hat{W}_1,
\end{equation}
where $\hat{F_\phi}$ is a diagonal matrix which applies a phase shift $e^{i\phi}$ in port $i$ (the other elements are independent of $\phi$), while $\hat{W}_1$ and $\hat{W}_2$ are the parts that precede and follow the phase shifter, respectively (Fig. \ref{fig:W_opt}(b)). Thus, Eq. (\ref{eqn:L}) becomes 
\begin{align}
\mathcal{R}\left\{ \mathbf{Y}^T \frac{d \hat{W}}{d\phi} \mathbf{X} \right\} &= \mathcal{R}\left\{ \mathbf{Y}^T \hat{W}_2 \frac{d\hat{F}_\phi}{d \phi} \hat{W}_1 \mathbf{X} \right\} \label{eqn:XY}
\\ \nonumber
& = -\mathcal{I}\left\{(\hat{W}_2^T \mathbf{Y})_i e^{i\phi} (\hat{W}_1 \mathbf{X})_i \right\},
\end{align}
where $(\mathbf{V})_i$ is the $i$-th element of the vector $\mathbf{V}$, and $\mathcal{I}$ denotes the imaginary part. This result can be written more intuitively in a notation similar to the main text. Namely, if $A_\phi$ is the field amplitude generated by input $\mathbf{X}$ from the right, measured right after the phase shifter corresponding to $\phi$, while $A^{\mathrm{adj}}_\phi$ is the field amplitude generated by input $\mathbf{Y}$ from the right, measured at the same point, then
\begin{equation}
\frac{d\mathcal{L}}{d\phi} = -
\mathcal{I}\left\{ A_\phi A^{\mathrm{adj}}_\phi \right\}
\end{equation}
By recording the amplitudes in all ports during the forward and the backward propagation, we can thus compute in parallel the gradient with respect to every phase shifter. Notice that, within this computational model, we do not need to go through the full procedure outlined in Section 4 of the main text. However, this procedure is crucial for the \textit{in situ} measurement of the gradients, and works even in cases which cannot be correctly captured by the simplified matrix model used here. 

\section{Training demonstration}

In the main text we show how the \textit{in-situ} backpropagation method may be used to train a simple XOR network.  Here we demonstrate training on a more complex problem. Specifically, we generate a set of one thousand training examples represented by input and target $(\mathbf{X}_0 \to \mathbf{T})$ pairs. Here, $\mathbf{X}_0 = [x_1, x_2, P, 0]^T$ where $x_1$ and $x_2$ are the independent inputs, which we constrain to be real for simplicity, and $P(x_1, x_2)= \sqrt{P_0-x_1^2-x_2^2}$ represents a mode added to the third port to make the norm of $\mathbf{X}_0$ the same for each training example.  In this case, we choose $P_0 = 10$.  Each training example has a corresponding label, $y \in \{0,1\}$ which is encoded in the desired output, $\mathbf{T}$, as $[1, 0, 0, 0].^T$ and $[0, 1, 0, 0].^T$ for $y = 0$ and $y=1$ respectively.

For a given $x_1$ and $x_2$, we define $r$ and $\phi$ as the magnitude and phase of the vector $(x_1, x_2)$ in the 2D-plane, respectively. To generate the corresponding class label, we first generate a uniform random variable between 0 and 1, labeled $\mathcal{U}$, and then set $y=1$ if
\begin{equation}
\exp{\left(-\frac{(r-r_0-\Delta \sin(2\phi))^2}{2\sigma^2}\right)} + 0.1~\mathcal{U} > 0.5.
\end{equation}
Otherwise, we set $y=0$.  For the demonstration, $r_0 = 0.6$, $\Delta = 0.15$, and $\sigma = 0.2$. The underlying distribution thus resembles an oblong ring centered around $x_1 = x_2 = 0$, with added noise. 

As diagrammed in Fig. \ref{fig:demo}(a), we use a network architecture consisting of six $4 \times 4$ layers of unitary OIUs, with an element-wise activation $f(z) = |z|$ after each unitary transformation except for the last in the series, which has an activation of $f(z) = |z|^2$. After the final activation, we apply an additional `softmax' activation, which gives a normalized probability distribution corresponding to the predicted class of $\mathbf{X}_0$. Specifically, these are given by $s(z_i) = \exp{(z_i)}/\left( \sum_j \exp{(z_j)} \right)$, where $z_{i = 1, 2}$ is the first/second element of the output vector of the last activation (the other two elements are ignored). The ANN prediction for the input $\mathbf{X}_0$ is set as the larger one of these two outputs, while the total cost function is defined in the cross-entropy form
\begin{equation}
\mathcal{\mathcal{L}} = \frac{1}{M}\sum_{m=1}^M \mathcal{L}^{(m)} = \frac{1}{M}\sum_{m=1}^M -\log(s(z_{m, t})),
\label{eq:cross_entropy}
\end{equation}
where $\mathcal{L}^{(m)}$ is the cost function of the $m$-th example, the summation is over all training examples, and $z_{m, t}$ is the output from the target port, $t$, as defined by the target output $\mathbf{T}^{(m)}$ of the $m$-th example. We randomly split our generated examples into a training set containing 75\% of the originally generated training examples, while the remaining 25\% are used as a test set to evaluate the performance of our network on unseen examples.

\begin{figure}
\includegraphics[width=\columnwidth]{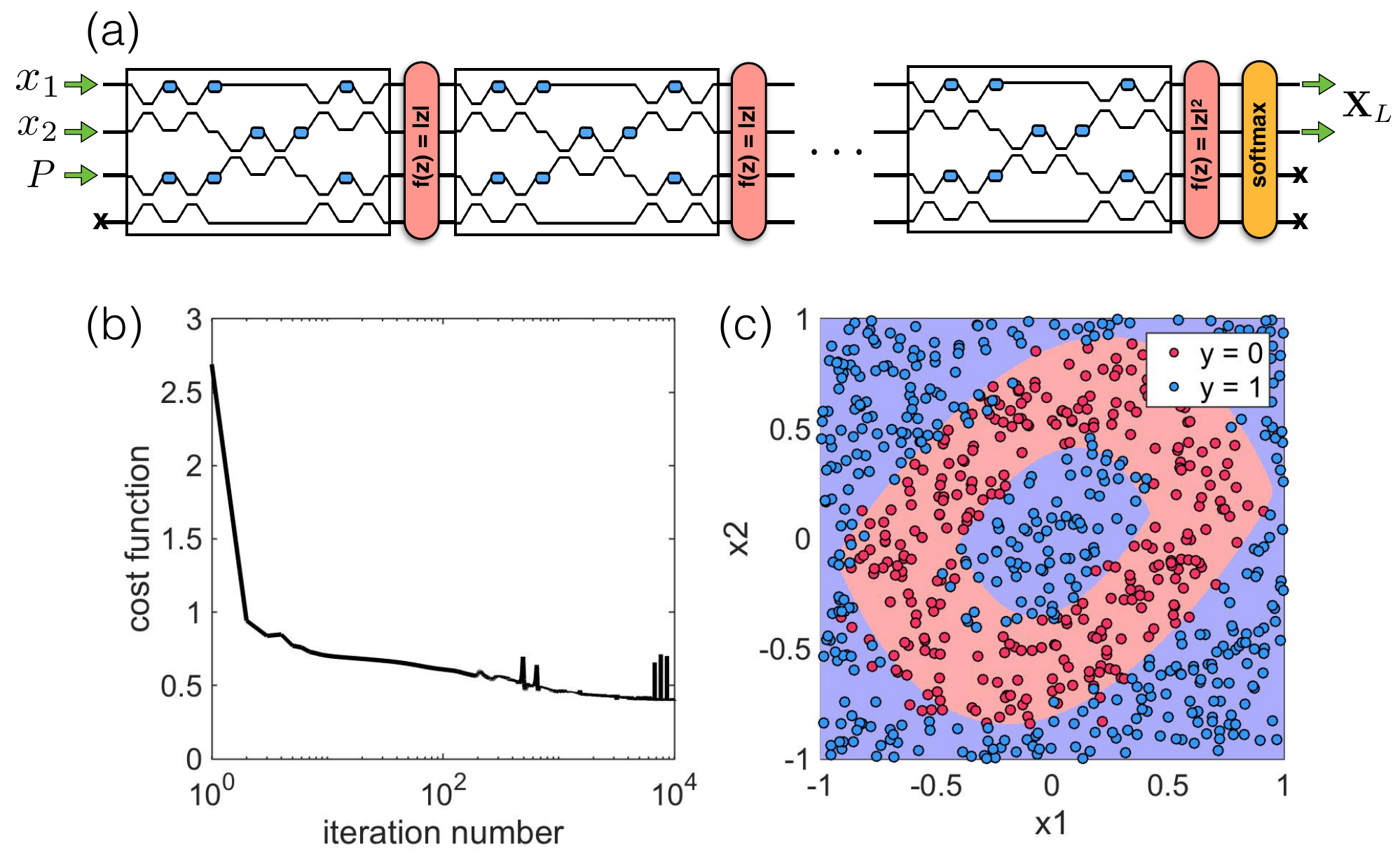}
\caption{Numerical demonstration of a photonic ANN learning to classify an oblong ring. (a) The architecture of the ANN.  Six layers of $4 \times 4$ OIUs with $|z|$ activations.  A final softmax activation is applied at the very end.  (b) The loss function of Eq. (\ref{eq:cross_entropy}) over training iterations. (c) The training examples, blue and red dots correspond to $y=0$ and $y=1$ labels on a given $x_1$ and $x_2$ input.  The background shows the prediction of the network on a continuum of $x_1$ and $x_2$ pairs, with colors representing the corresponding predictions.  One can see that the ring was learned successfully without overfitting.  \label{fig:demo} }
\end{figure}

As in the XOR demonstration, we utilized our matrix model of the system described in Section \ref{pnn}. As in the main text, at each iteration of training we compute the gradient of the cost function with respect to the phases of each of the integrated phase shifters, and sum this over each of the training examples. For the backpropagation through the activation functions, since $|z|$ and $|z|^2$ are non-holomorphic, we use eq. \ref{eqn:noholo} from Section \ref{noholo}, to obtain
\begin{align}
\bm{\delta}_L &= 2 \mathbf{Z}_L^* \odot \mathcal{R}\{\bm{\Gamma}_L\} \\
\bm{\delta}_l &= \exp(-i\bm{\phi}_l) \odot \mathcal{R}\{\bm{\Gamma}_l\},
\end{align}
where $\bm{\phi}_l$ is a vector containing the phases of $\mathbf{Z}_l$ and $\bm{\Gamma}_L$ is given by the derivative of the cross-entropy loss function for a single training example 
\begin{equation}
\bm{\Gamma}_L = \frac{\partial \mathcal{L}^{(m)}}{\partial z_{m, i}} = s(z_{m, i}) - \delta_{i, t},  
\end{equation}
where $\delta_{i, t}$ is the Kronecker delta. 

With this, we can now compute the gradient of the loss function of eq. \ref{eq:cross_entropy} with respect to all trainable parameters, and perform a parallel, steepest-descent update to the phase shifters, in accordance with the gradient information. Our network successfully learned the this task in around 4000 iterations. The results of the training are shown in Fig. \ref{fig:demo}(b). We achieved a training and test accuracy of 91\% on both the training and test sets, indicating that the network was not overfitting to the dataset.  This can also be confirmed visually from Fig. \ref{fig:demo}(c).  The lack of perfect predictions is likely due to the inclusion of noise.

\end{document}